\documentclass{article}
\usepackage[utf8]{inputenc}
\usepackage[textwidth=0.75in]{todonotes}
\usepackage{wrapfig}
\usepackage{tabularx}
\usepackage{graphicx}
\usepackage[export]{adjustbox}
\usepackage{hyperref}

\usepackage[top=1in,bottom=1in,inner=1in,outer=1in,marginpar=2in]{geometry}

\title{Managing, Analyzing and Sharing Research Data \protect\\
with Gen3 Data Commons}
\author{Craig Barnes, \and  Kyle Burton, \and Michael S. Fitzsimons, \and Hara Prasad Juvvala, \and Brienna Larrick, \and Ao Liu, \and Clint Malson, \and  Noah Metoki-Shlubsky, \and
Christopher Meyer, \and Andrii Prokhorenkov, \and Jawad Qureshi, \and Radhika Reddy, \and Pauline Ribeyre, \and L. Philip Schumm, \and Mingfei Shao, \and Trevar Simmons, \and Alexander VanTol, \and Peter Vassilatos, \and Aarti Venkat, \and Robert L. Grossman  \\
\protect\\
Center for Translational Data Science \\
University of Chicago
}
\date{July 14, 2025}

\begin{document}

\maketitle

\tableofcontents

\clearpage

\section*{Abstract}

Gen3 is an open-source data platform for building data commons.   A data commons is a cloud-based data platform for managing, analyzing, and sharing data with a research community.  Gen3 has been used to build over a dozen data commons that in aggregate contain over 28 PB of data and 64 million FAIR data objects.  To set up a Gen3 data commons, you first define a data model.  Gen3 then autogenerates 1) a data portal for searching and exploring data in the commons; 2) a data portal for submitting data to the commons; and 3) FAIR APIs for accessing the data programmatically.  Gen3 is built over a small number of standards-based software services, which are designed to support current and future Gen3 components so that Gen3 can interoperate with other data platforms and data ecosystems.

\clearpage
\addcontentsline{toc}{section}{Main}
\section*{\hbox{}}

Despite the well understood importance and benefits of sharing research data, it is still a challenge for most research projects and communities to share the data that their researchers generate in a way that is integrated with the tools, applications, and software services used by their community.  A {\em data commons} is a cloud-based data platform with a governance structure to manage, analyze and share data to support a research community \cite{grossman2023ten}.  Data commons generally clean and curate submitted data, align it with a data model, and sometimes harmonize it by processing it with a common set of workflows.   A data commons also supports the interactive exploration of its data by integrating the analysis and visualization tools and services that its community uses.

In this article, we describe Gen3, which is an open-source data platform for building and operating data commons \cite{gen32024github}.  Gen3 has been used to develop and operate over 15 data commons, which in aggregate, make over 28 PB of data available to the research community, spanning more than 64 million FAIR data objects.

Gen3 has been used as part of the infrastructure for large-scale data commons, such as the NIH National Cancer Institute (NCI) Cancer Research Data Commons (CRDC) \cite{brady2024nci, wang2024nci}, the NIH National Heart, Lung, and Blood Institute (NHLBI) BioData Catalyst data platform \cite{ahalt2023building}, and the NIH National Institute of Biomedical Imaging and Bioengineering Medical Imaging and Data Resource Center (MIDRC) \cite{baughan2022sequestration}.

\begin{figure}[ht]
    \includegraphics[width=1.0\textwidth]{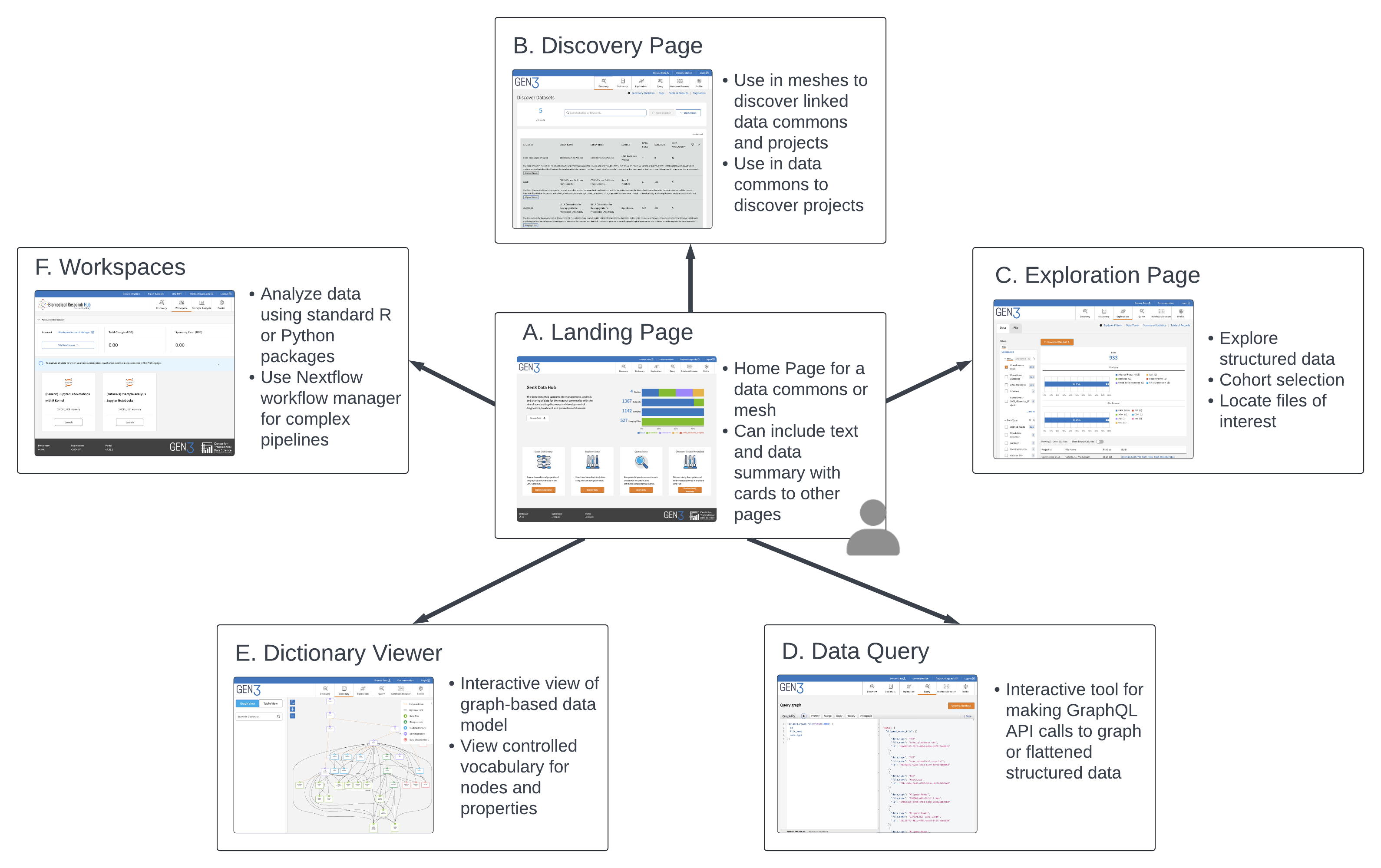}
        \caption{Overview of how a user interacts with the Gen3 Data Portal.  A) Users begin on a landing page that summarizes the data content; B) The Discovery page provides discoverability of project level summary data;  C) The Exploration Page allows users to explore data, build cohorts or create subsets of data, and find files of interest based on their selection criteria; D) Users can programmatically query data on the Query page or through an API query; E) The Dictionary Viewer allows users to explore the types of data present in the system and helps data submitters understand how to structure data for submission; and F) Workspaces leveraging both commercial (e.g. Stata) and non-commercial (e.g. R or Python) software can be used to analyze data. Note that not all of these services need be deployed for a given system.}
    \label{fig:gen3_front_end}
\end{figure}

Importantly, Gen3 has also been used  to set up smaller scale data commons to support particular research communities, such as the liquid biopsy research community \cite{grossman2021bloodpac}, the Connecting Health Outcomes Research Data Systems (CHORDS) research community (\url{chordshealth.org/discovery}), and a data commons to support COVID-19 research during the pandemic for the Chicago region \cite{trunnell2024pandemic}, which is now focused on the study of Long COVID.  See Figure~\ref{fig:gen3_front_end} for an example of a Gen3 data commons from a user perspective.

Gen3 supports cloud-based data objects, structured data, and semi-structured data. All three types of data are managed by Gen3 in a FAIR manner \cite{wilkinson2016fair}.  High-level features of a Gen3 system include:

\begin{enumerate}

\item {\bf Native FAIR \cite{wilkinson2016fair} support for structured, semi-structured, and unstructured data.}  Gen3 provides APIs so that structured data from databases, semi-structured JSON-based data, and unstructured data objects are all managed as findable, accessible, interoperable, and reusable (FAIR) data \cite{wilkinson2016fair}.

\item {\bf Data model driven architecture.} Gen3 is a data-model driven architecture in the sense that once a data commons administrator defines a YAML-based data model, the Gen3 software auto-generates FAIR APIs  \cite{wilkinson2016fair} for the data, as well as a web-based portal for discovering, exploring data, and submitting data to the commons.  Whenever the data model is updated, the FAIR APIs and data portals are auto-generated again. This approach reduces the effort required to build data commons and keep them up to date as their data changes over time. This is especially important for smaller projects, which often do not have a dedicated data scientist for data curation and ingestion.

\item {\bf Cloud-native with elastic scalability.} Gen3 is cloud native and uses the elastic scale-out properties of cloud computing infrastructure to manage petabytes or tens of petabytes of genomic, imaging, and other large-scale biomedical data.

\item {\bf Narrow middle architecture.} The design of Gen3 is based upon the end-to-end system design principle \cite{saltzer1984end} that is the architectural basis of the internet.  An end-to-end distributed system design for data platforms is based upon a small number of distributed software services that are used to support applications that produce data for the system at one end and consume data for the system at the other end \cite{grossman2018narrow-middle}. With this approach, it is easier to add new sources of data (at one end) and new applications for working with data (at the other end) over time, without changing the core architecture of the system.

\item {\bf Life cycle support for data.} Gen3 provides life cycle support for projects. All data objects managed by Gen3 data commons are assigned persistent digital IDs compliant with the GA4GH DRS data standard \cite{rehm2021ga4gh}.  Gen3 persistent identifiers are opaque and do not specify the physical location of the data in a particular cloud, making it easy to move the data from one cloud-based location to another. Also, Gen3 structured data can be exported in a self-describing, encapsulated format \cite{lukowski2022PFB} that can be versioned, assigned a persistent Gen3 identifier, and then imported at a later time into another data commons.
\end{enumerate}

In the remainder of the article, we describe the key features and overall architecture of Gen3 and how it is used to enable research and leverage the very large datasets that are now the norm in most branches of science.

\addcontentsline{toc}{section}{Results}

\section*{Results}

\addcontentsline{toc}{subsection}{Gen3 Data Commons}
\subsection*{Gen3 Data Commons}
\begin{figure}
    \includegraphics[width=0.75\textwidth, center]{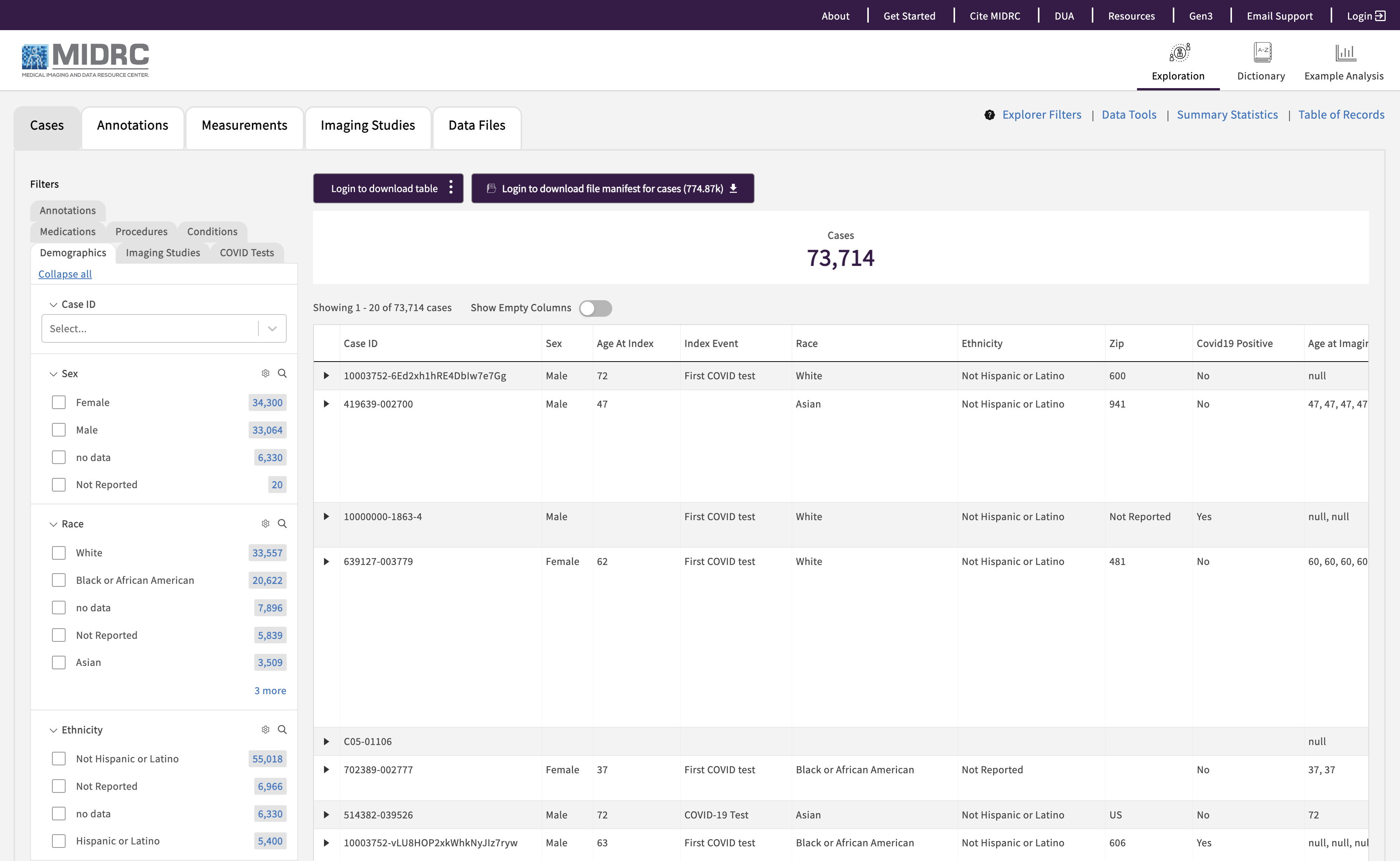}
        \caption{The Exploration portal is the main entry point for users to query and explore data found in a Gen3 data commons.  An example from the MIDRC Data Portal is shown here \cite{midrc}.  Facets, tabs, and charts are all customizable for a particular data commons.}
    \label{fig:explorer}
\end{figure}

Gen3 is currently used by over a dozen different organizations around the world to develop and operate data commons. Organizations that operate Gen3 data commons include universities, academic medical centers, industry, and government agencies.
As an example, the Center for Translational Data Science, the primary developer and maintainer of Gen3, operates ten data commons that, in aggregate, make over 64 million FAIR data objects, containing information about over 2 million research subjects, spanning over 28 PB of data, available to the research community.  Each organization that operates a Gen3 data commons can configure Gen3 to support their policies for sharing data and can customize the Gen3 portals and add data analysis tools to support their research community. The Gen3 Community (gen3.org/community/) includes a Slack channel, a discussion board, a help desk, and regular virtual community meetings. Figure~\ref{fig:explorer} shows a typical Gen3 portal for searching for data and creating cohorts.

Gen3 has been primarily utilized for biomedical data, although it is fundamentally field agnostic.  A few significant examples of Gen3 data commons include: NHLBI BioData Catalyst \cite{BDC}, Data Commons Framework, which is part of the NCI Cancer Research Data Commons \cite{brady2024nci}, NIBIB MIDRC Data Commons \cite{midrc}, the Kids First Data Resource Portal \cite{kidsfirst}, and the Veterans Precision Oncology Data Commons \cite{Elbers2020-nv}. Gen3 provides a data platform so that organizations can build data commons broadly similar to the NCI Genomic Data Commons \cite{heath2021nci}.

\begin{figure}
    \includegraphics[width=1.0\textwidth]{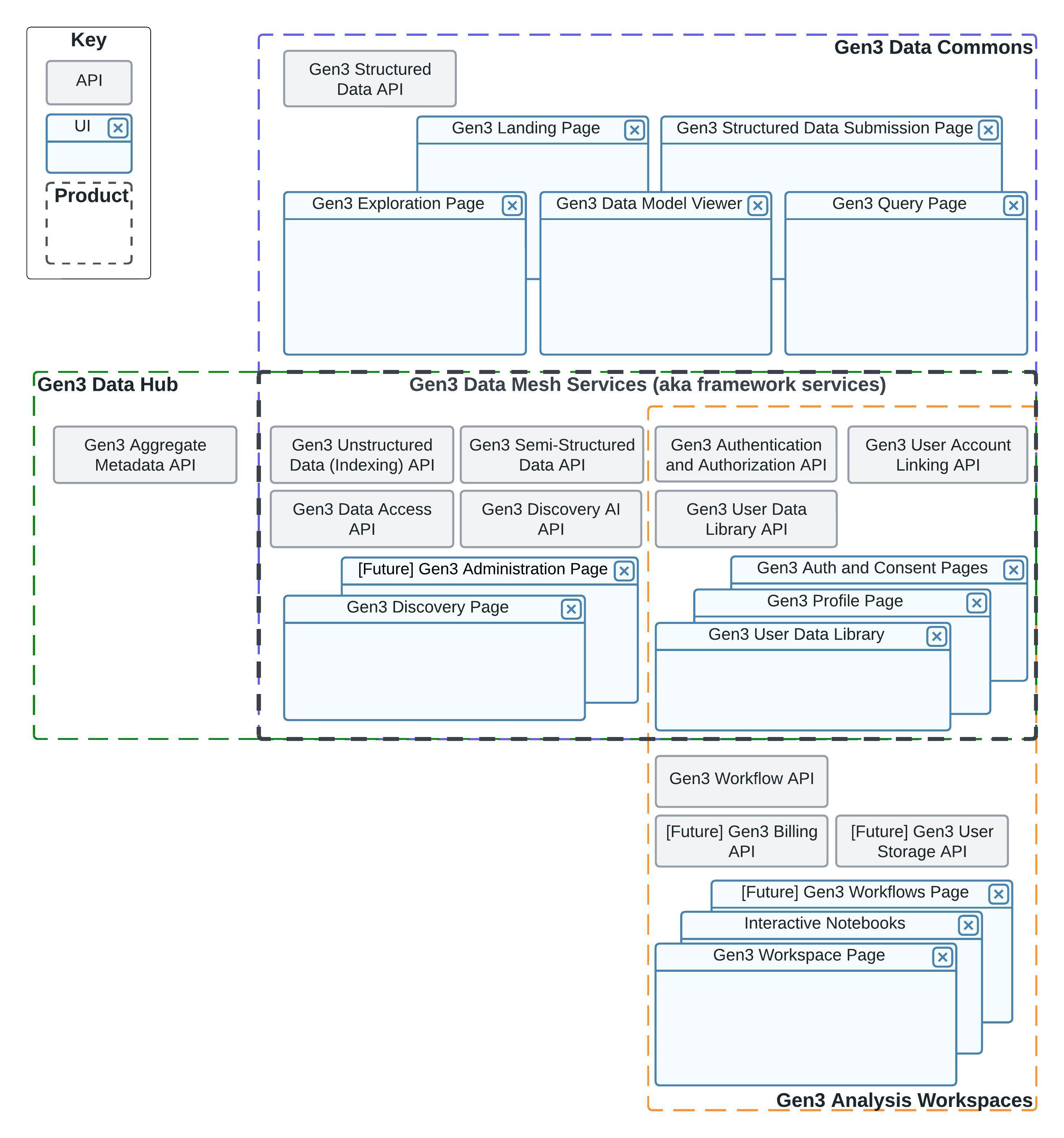}
        \caption{The different Gen3 products include Data Commons, Data Hub, Data Mesh Services, and the Analysis Workspace.  The figure depicts which APIs and UIs are included for each product and how the products relate to each other.}
    \label{fig:gen3_products}
\end{figure}

Gen3 is a modular, open-source software platform that exposes a standard set of application programming interfaces (APIs) and user interfaces (UIs). It is capable of managing various types of cloud-based data and enabling cloud-based compute over those data. The underlying software and APIs are designed from the ground up to be interoperable, standards-based, and configurable.  The Gen3 software is cloud agnostic and its deployment is facilitated by containerization and orchestration frameworks such as Kubernetes.  Figure~\ref{fig:gen3_front_end} shows some of the main Gen3 portal components, which include an Exploration portal for creating cohorts or subsets of data for further analysis, a Discovery portal for searching for datasets of interest,  workspaces for analyzing data, a dictionary viewer for exploring the data model, and a query builder for interactively creating GraphQL queries.  The supplementary material contain additional information about Gen3 components (Figure~\ref{fig:gen3_products}).

\addcontentsline{toc}{subsection}{Data Model Driven Architecture}
\subsection*{Data Model Driven Architecture}

\begin{figure}[ht!]
    \includegraphics[width=1.0\textwidth]{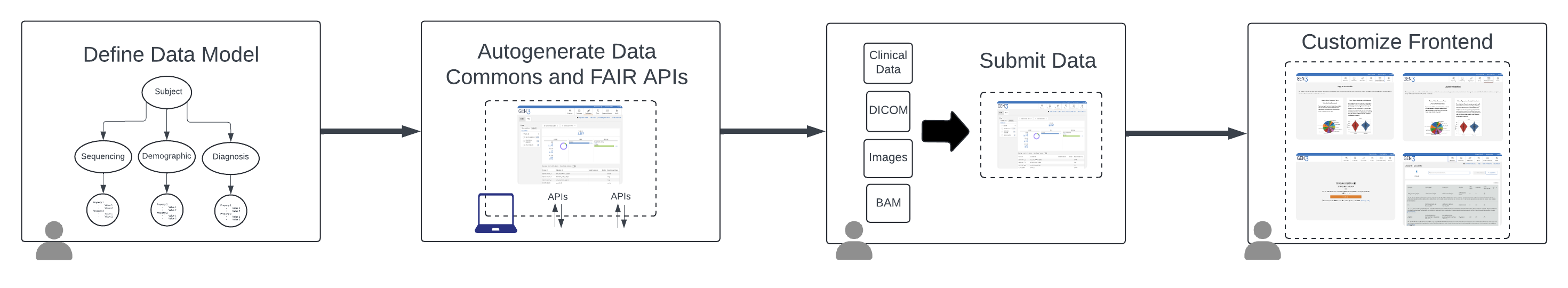}
        \caption{Stages of bringing up a Gen3 Data Commons.  In the first step, a data commons administrator defines a data model.  Next, the system autogenerates a Data Commons and FAIR APIs. Then the operator submits data files and structured data.  Finally, the frontend is customized to adapt to user needs.}
    \label{fig:gen3_stages}
\end{figure}

An important design principle of the Gen3 data platform is that it is a data model driven architecture.  {\em Once a data model is specified, FAIR APIs are generated for the data and data portals for exploring and submitting data are  generated automatically over these APIs.}  In other words, a core data commons is created automatically once the data model is specified.  The data portals for exploring and submitting data can then be further customized as required. See Figure~\ref{fig:gen3_stages}.

The Gen3 data model is a graph data model, with nodes representing entities and edges representing relationships between entities.  Properties are associated with each entity.   This is the same type of data model that is used by the NCI Genomic Data Commons \cite{heath2021nci}. As mentioned above, once the Gen3 data model is defined, the FAIR APIs and portals using these APIs are automatically generated.

Nodes are used to define a group of related terms or properties.  Some examples could include: patients, sequencing experiments, assays, clinical visits, etc.  Clinical variables like a cancer diagnosis or a subject’s sex might go into the diagnosis or demographic nodes, respectively. Variables related to how a biological sample was collected or processed may be found in a biospecimen node. Data files, such as medical images or genomic files, can also be nodes and have associated metadata variables like file size, format, and file name. The relationships or connectivity between nodes can also be informative.  For example, a sample node may derive from or connect to a specific patient, or a genomic data file may also be associated with a specific sample.

Properties are assigned different types including: string, boolean, floating point number, integer, or enumeration. Properties can also be defined as arrays of any of these types. The acceptable values for properties can be further restrained by defining regex patterns that strings must match or minimum or maximum values for numeric data. Nodes, properties, and permissible values are specified in a series of YAML files.

\addcontentsline{toc}{subsection}{FAIR Data, Data Mesh Services and the Narrow Middle Architecture}
\subsection*{FAIR Data, Data Mesh Services and the Narrow Middle Architecture}

Gen3 Data Mesh Services (aka Framework Services) are a minimal set of distributed software services that provide open APIs that form the foundation, or ``framework'', for building systems. Such foundational support includes indexing data objects (both unstructured and semi-structured), associating metadata with the data objects, controlling user access to data via a policy engine, and providing FAIR APIs for structured data.  Gen3 data mesh services can be deployed alone to provide FAIR APIs for data in a cloud bucket, for example.  It is more common, though, that they are deployed as part of a broader product like a Gen3 Data Commons.

Whenever possible, Gen3 uses existing standards and solutions for its data mesh services. In particular, Gen3 Data Mesh Services follow well-accepted standards, including OpenID Connect, OAuth 2.0, and GA4GH DRS \cite{rehm2021ga4gh}.

The design of Gen3 Data Commons around these core services is an example of the end-to-end design principle \cite{saltzer1984end}.  Within the data commons community, this is usually called the {\em narrow middle architecture} \cite{grossman2018narrow-middle}, since at one end of the framework different applications use the core services to clean, curate and ingest data, while at the other end of the framework, other applications use the core services to process the data, analyze the data, and share the data.  As an illustration of the importance of basing the Gen3 data platform on this architectural principle, when the first Gen3 data commons were set up during the period 2016-2018, Jupyter Notebooks were not common, but as Jupyter Notebooks grew in popularity, it was easy to add them as another supported application on the consumer side of the narrow middle architecture.

With Gen3 Data Mesh Services, the data objects, structured data, and semi-structured data supported by Gen3 can all be made Findable, Accessible, Interoperable, and Reusable (FAIR) \cite{wilkinson2016fair}. As just described, Gen3 Data Mesh Services are used to assign persistent identifiers to data, to manage metadata associated with these identifiers, and to support authentication, authorization and policy-based access to registered and controlled access/sensitive data.

Common examples of biomedical data objects include genomic BAM or CRAM file, images, sensor data, bulk clinical and phenotypic data, and videos.  These are typically handled within Gen3 as unstructured cloud data objects. A {\em cloud data object}, or more simply a {\em data object}, is unstructured data with an associated persistent identifier and associated metadata. As an example, data objects in AWS are managed using the S3 service.

In Gen3, structured data are data that adhere to a specific and strict schema, generally a graph data model, though Gen3 data commons can be built using other data models, such as an OMOP, relational, or FHIR data model.

The data model is typically used to represent a harmonized version of the data.  Gen3 also uses a data dictionary so that structured data can be cleaned, curated, and aligned with respect to the data dictionary.

Semi-structured data are organized as unique identifiers with flexible key/value pairs (including nesting). The key/value pairs may be consistent between records, but are not required to be. This is typically used for storing publicly available metadata about available datasets or additional public metadata about samples.

Gen3 sometimes uses the term {\em data hybrid architecture} to refer to the fact that it supports data objects, semi-structured data, and structured data.

\addcontentsline{toc}{subsection}{Cloud Native and Scalable}
\subsection*{Cloud Native and Scalable}

Gen3 is cloud native, cloud agnostic, and scalable so that it can support a wide range of different research groups and different research communities.  In particular, Gen3 data commons in production support a wide range of data types and data sizes, from TBs to tens of PBs of data. In more detail, Gen3 can be run on the main commercial cloud platforms (AWS, GCP and Azure) as well as on-premise OpenStack software. It was designed from its beginnings to be cloud native and leverages the natural horizontal scalability of cloud architectures.

Gen3 services and applications are all containerized and Gen3 is designed to run over Kubernetes (K8s), which is an open-source system to automate the deployment, scaling, and management of containerized applications.  Gen3 uses Karpenter, which is an open source Kubernetes cluster autoscaler.  This allows a project to easily transition as the number of its users grows (or shrinks) over time.  More recently, Gen3 has transitioned to Helm charts, which is a Kubernetes package manager, and has significantly decreased the time required for Gen3 system administrators to quickly deploy a new Gen3 system.

\addcontentsline{toc}{subsection}{Life Cycle Support for Data}
\subsection*{Life Cycle Support for Data}

\begin{figure}[ht!]
\centering
\includegraphics[scale=0.65]{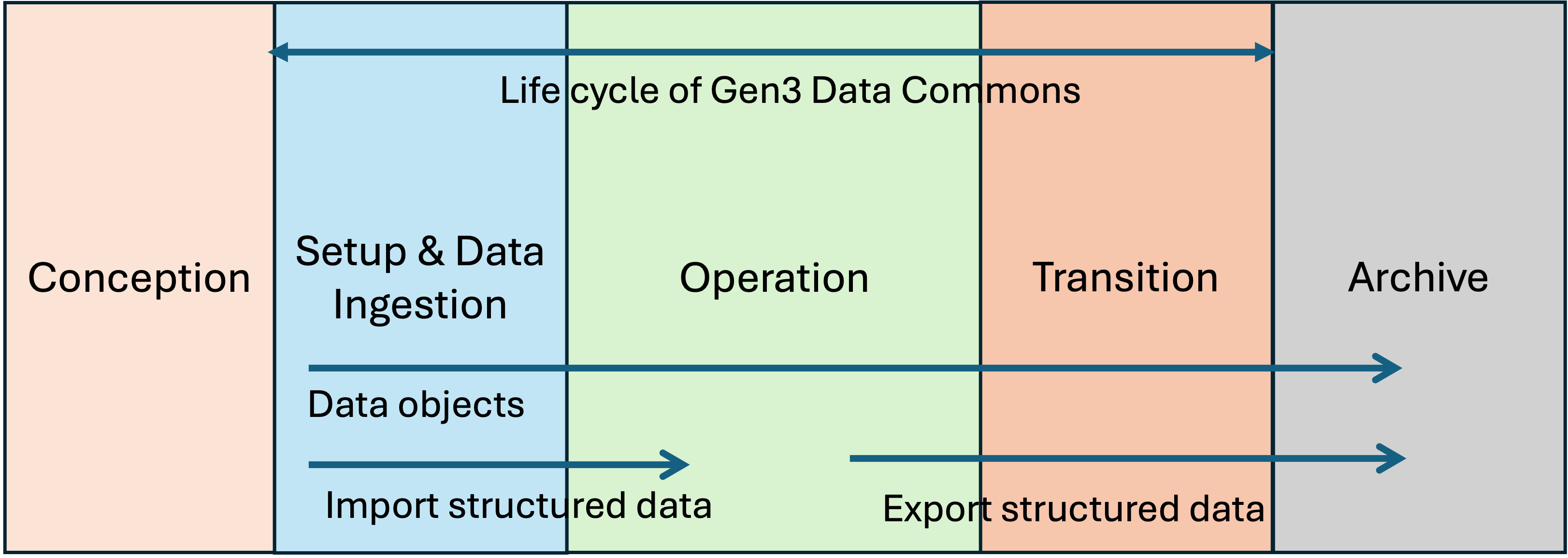}
\caption{Gen3's support for the full life cycle of the data it manages.}
\label{fig:life-cycle}
\end{figure}

Some data commons support projects with fixed known duration, such as five or ten years.  To support these types of projects, and, more generally, data commons that must be retired after a period of time, Gen3 provides functionality to preserve and archive data, even beyond the life of a project. As Figure~\ref{fig:life-cycle} illustrates, a project can set up a Gen3 data commons for its lifetime --- say 3, 5 or 10 years.  At the end of this period,  1) all or selected data objects can be migrated to a persistent cloud-based archive by simply updating their DRS persistent identifiers \cite{rehm2021ga4gh}; and, 2) the structured data can be i) exported and imported into other commons or repository or ii) packaged in a bulk format (such as Avro \cite{lukowski2022PFB} or bulk-FHIR \cite{jones2021bulk-fhir}) and archived as a data object in a data repository.  Importantly, an advantage of this approach is that projects can have a customized data commons to accelerate their research during the lifetime of the project, and since their commons use the Gen3 framework services it is relatively straightforward to migrate the data objects and structured data to a long-term repository after the project is over, always keeping the data FAIR.

\addcontentsline{toc}{subsection}{Gen3 Components}
\subsection*{Gen3 Components}

\subsubsection*{Gen3 Data Commons}

\begin{figure}[ht]
    \includegraphics[height=0.91\textheight]{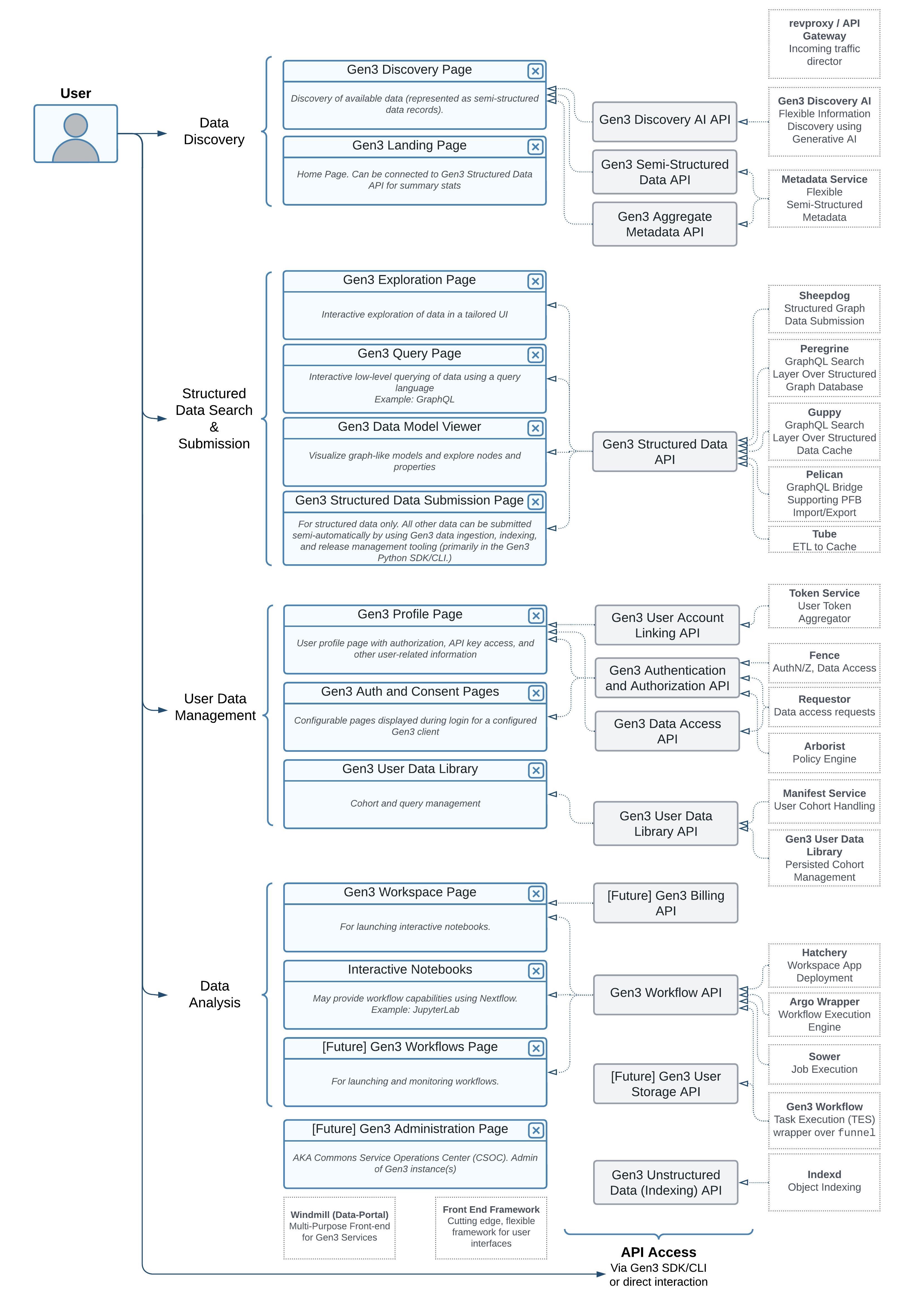}
    \caption{Overview of how a user interacts with Gen3. Gen3 user interface components are in blue and the Gen3 API components that support the user interface components are in gray.}
    \label{fig:product_arch}
\end{figure}

As Figure~\ref{fig:product_arch} shows, a Gen3 data commons consists of web portals for exploring data via its attributes and building virtual cohorts (``Exploration Page''), for discovering datasets (``Discovery Page''), for submitting data (``Data Submission Page''), for managing Jupyter Notebooks (``Workspace Page''), and for various other administrative activities.

Each of these pages may be customized.  In particular, Gen3 includes a Frontend Framework for adding software applications and tools to the Exploration, Discovery and other Gen3 pages.  The Gen3 Frontend Framework consists of two primary modules --- the Core Module and the Frontend Module --- and a NextJS web application, which together create the Gen3 data commons user interface (UI).  This architecture reduces code complexity, abstracts UI interactions with Gen3 services, supports customization, and simplifies deployment and cost.

The Core Module interfaces with Gen3 services and analysis tools for managing user context. This includes the current cohort, selected studies, files, and active analysis tools. The Frontend Module provides Gen3 services and API access via ``hooks,'' allowing seamless data reading and updates. This design isolates pages and components from Gen3 API changes and standardizes service access. Additionally, the Core Module includes the Analysis Tool Framework (ATF), which supports the development of applications using Gen3 services and third-party APIs. The ATF maintains a registry of tools, along with metadata describing their applicable contexts.

Analysis tools are developed using React, Gen3 core/frontend packages, and custom code. These tools are published as NPM modules and registered as plugins with the ATF during the server build, making them accessible via the data commons’ root URL.

\subsubsection*{Gen3 Analysis Workspaces}

Gen3 Analysis Workspaces (``Gen3 Workspace'') provide secure, cloud-based data management and analysis environments. These are typically fully integrated within a specific data commons but can also be set up as standalone analysis environments. Users may access data directly from the hosting commons according to their permissions, or may access data from other cloud or web-based resources (e.g., other Gen3 commons or data repositories) using the Gen3 data mesh services to authenticate to the remote resource (if necessary) and retrieve data. Data may also be uploaded into a workspace. Cloud compute costs may be covered by the hosting commons or may be paid by the user (e.g., via an NIH STRIDES account or credit card), as appropriate.

Gen3 Workspaces currently utilize either a JupyterLab or RStudio interface, though any environment that provides a web-based interface could also be used. JupyterLab kernels provide access to a variety of commercial and free software including Stata, R, Python, Julia, etc. Gen3 Workspaces provide persistent, user-specific file storage. Specifically configured workspaces are available for creating and executing workflows (e.g., using Nextflow). Long-running workflows or pipelines may be executed using the Gen3 Workflow Execution API.

By default, Gen3 workspaces are flexible and open, allowing users to install their own software (with limitations), customize their environment using standard Linux utilities, utilize other web-based services for collaboration and sharing (e.g., GitHub), and upload and download their own files. Each of these may be disabled, if necessary, to create a more controlled environment (e.g., a data enclave). This inherent flexibility with selective hardening means that Gen3 workspaces are more powerful, and can be used to provide a better user experience, than typical workspaces attached to a data repository or within a data enclave.

\addcontentsline{toc}{section}{Discussion}
\section*{Discussion}

As explained in \cite{grossman2024ten}, data commons are usually understood to differ from data repositories in several ways, including by curating and harmonizing the data submitted to it and by integrating applications, tools, and software services for exploring and analyzing the data contained in the commons.    Data commons are also developed from the ground up to be FAIR, unlike many data repositories, which rely more on portal based access to data.

Interoperating two or more data commons or a data commons with other cloud-based computational or other resources is of increasing interest, and Gen3 provides the FAIR interfaces and other APIs to support this type of interoperability \cite{grossman2024framework, grossman2024ten}.  In addition, Gen3 data mesh services can be used to support the interoperability of multiple cloud-based resources, whether they are Gen3-based or not.

Gen3 data commons are designed to be scalable and secure cloud-based data platforms that can support workflows to clean, curate and harmonize data. Given these capabilities, human resources with different skill sets and expertise are usually needed to set up and operate a Gen3 data commons.  Broadly speaking, this expertise splits naturally into several categories: 1) expertise about Gen3 commons per se; 2) cloud computing expertise; 3) software development expertise; 4) data expertise; and, 5) security and compliance expertise.

Sometimes the term commons services provider is used for the skills in \#1 above to set up, configure, operate, and maintain a Gen3 data commons and the associated Gen3 data mesh services.  To help with \#1, there is open-source software available in gen3.org to set up and operate what is sometimes called a Commons Services Operations Center (CSOC) that can be used to automate in part the setup and operation of one, two, or more Gen3 data commons.

With this terminology, a Gen3 data commons requires: 1) a {\em commons services provider},  who sets up, configures, and maintains the Gen3 data mesh services and other cloud-based software services required for a Gen3 commons; 2) a {\em commons administrator}, who configures a particular commons, sets up and updates the data model, and performs related services;  and 3) {\em software developers and data scientists}, who customize the commons, ingest, curate and harmonize submitted data, and perform related functions. Additional information about what is required to operate a data commons is provided in the Methods Section below.

As explained in \cite{grossman2023ten}, the success of a commons often depends upon how well the data cleaning, curation and harmonization are done.  Successful commons often include a Data Coordinating Center or work closely with a Data Coordinating Center to provide these services. For example, the Genomic Data Commons \cite{heath2021nci} includes dedicated services for cleaning, curating, and harmonizing submitted data.

\clearpage

\addcontentsline{toc}{section}{Methods}
\section*{Methods}

In the following sections, we provide an overview of how to set up, customize, and operate a Gen3 data commons.  The key steps, in order, are: 1) define the data model; 2) deploy the system; 3) submit data; 4) customize the portal; and, 5) develop any custom applications required.   For more details on using, developing, and operating a Gen3 data commons, see the Gen3 documentation at https://docs.gen3.org.

\addcontentsline{toc}{subsection}{Define a Data Model}
\subsection*{Define a Gen3 Data Model}

A Gen3 data model is used to constrain the terminology used for describing patients, diagnoses, experimental procedures, and other details about the data in a data commons. Many Gen3 operators begin with a default basic dictionary we have created based on our experiences (https://github.com/uc-cdis/datadictionary/).  This can then be modified by adding or removing nodes, properties, and values.  Because harmonization can be a complex task, many data commons choose not to store all the structured data in the data model, but rather focus on those properties needed to create cohorts and leave other structured data to be stored as spreadsheets and downloaded for additional investigation.

The data model is specified in a YAML format where each node is defined in a separate schema file.  An example demographic node is shown in the Supplementary Figure \ref{fig:demographic.yaml}. A \textit{node ID} is used for data query/submission, \textit{category} is used to group nodes conceptually, the \textit{description} provides an overall description of the data associated with the node, and the list of links defines the \textit{relationship} to other nodes. Any properties marked as {\em required} must be included when submitting a node to the commons.

\addcontentsline{toc}{subsection}{System Deployment}
\subsection*{System deployment}

\begin{figure}[h!]
    \includegraphics[width=0.9\textwidth]{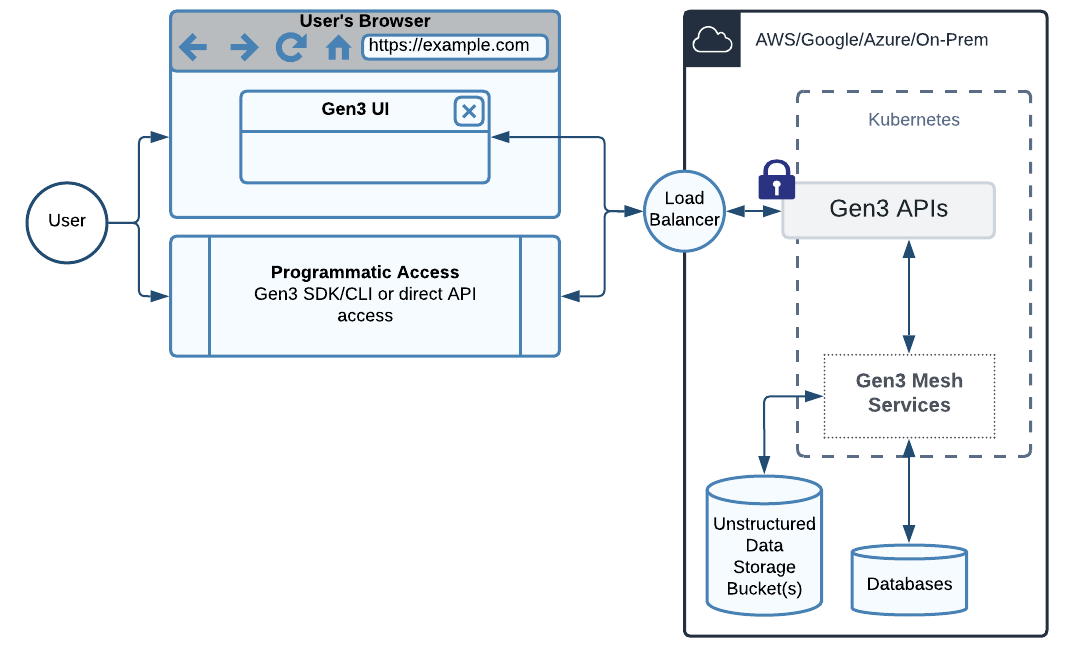}
        \caption{High level Gen3 Architecture}
    \label{fig:architecture}
\end{figure}

Gen3 is platform agnostic and can be run in any public cloud or private on-premise cloud. Figure~\ref{fig:architecture} shows a high-level view of the Gen3 system architecture.   Deployments of services and containerized workflows are managed using Kubernetes \cite{burns2022kubernetes},  which is an open-source system for automating deployment, scaling, and management of containerized applications.  Kubernetes has become the industry standard for managing cloud-native workloads.  It can currently be used across all cloud providers, including Amazon AWS, Google GCP, and Microsoft Azure, as well as for local or on-premise installations.

Gen3 relies upon Helm \cite{helm} to manage installation and management of Kubernetes applications.  Helm is used to build ``charts,'' which are packages of Kubernetes resources that are used to deploy apps to a cluster. The use of Helm has resulted in major improvements to the speed and ease with which Gen3 can be deployed.  Terraform is used to manage the cloud infrastructure necessary for running Gen3 within AWS.

Configuration instructions are found with the Gen3 commons manifest, which include portal configurations, ETL mappings, and versions for all services that will be deployed. Examples from CTDS-managed commons can be found at \url{https://github.com/uc-cdis/gen3-gitops}.

\addcontentsline{toc}{subsection}{Data Submission}
\subsection*{Data submission}

Gen3 accepts three main types of data: structured data (the type of information you would find in a database), unstructured data object files, and semi-structured JSON formatted data.  The ingestion of unstructured data in Gen3 includes uploading the file to cloud storage, minting a new GUID for the file, and indexing it along with the file's attributes such as file size, name, md5sum, and storage location. An associated record in the graph database can be created.  The file indexing service does not open files or consider their contents, and the only file metadata that is required to be accurate is the file size and the md5sum hash.

Structured data  are ingested into Gen3 via a service that checks conformation to the data model and then ingests the data into a database that manages the graph data. If invalid data are provided --- for example, a string value is provided for a numeric property --- the data submission will fail and the submission API will return an error message pointing out the mistakes.  If the metadata submission is valid, then those data are created or updated in the postgres database and are immediately accessible via the Gen3 query and data export services.

Finally, semi-structured data can be submitted to the metadata service (MDS) in Gen3. The Gen3 Python SDK tool has functions for making it easy to create discovery metadata or PPRL crosswalk metadata, both of which are powered by the MDS service.  JSON of any structure can be submitted to the MDS, which makes it a good option for including data from different projects that have no particular common format.  Common fields can then be made searchable (see data portal customization below).

\addcontentsline{toc}{subsection}{Customize Data Portal}
\subsection*{Customize Data Portal}\label{customize data portal}

There are many aspects to the data portal that can be customized by individual projects.  This includes cosmetic changes (e.g., changing the charts or text on the landing and Exploration pages, adding links to external documentation, or the use of different colors) as well as more substantive changes like making different data available for query and cohort building.

The Exploration page is the primary entry point for users within most data commons. To make data searchable in the Exploration page an ETL must be defined to bring data into an elasticsearch (ES) index from the postgres database where it was submitted.  Additional details can be seen in the Tube microservice documentation found on GitHub (https://github.com/uc-cdis/tube).  The advantage of ES queries is that they run much faster than postgres queries, which is important for data exploration.  You can see in Figure \ref{fig:explorer} an example of an Exploration page from the MIDRC project. The faceted search, tabs, and charts are all customizable for particular projects by modifying a configuration file.

\begin{figure}
    \includegraphics[width=0.75\textwidth, center]{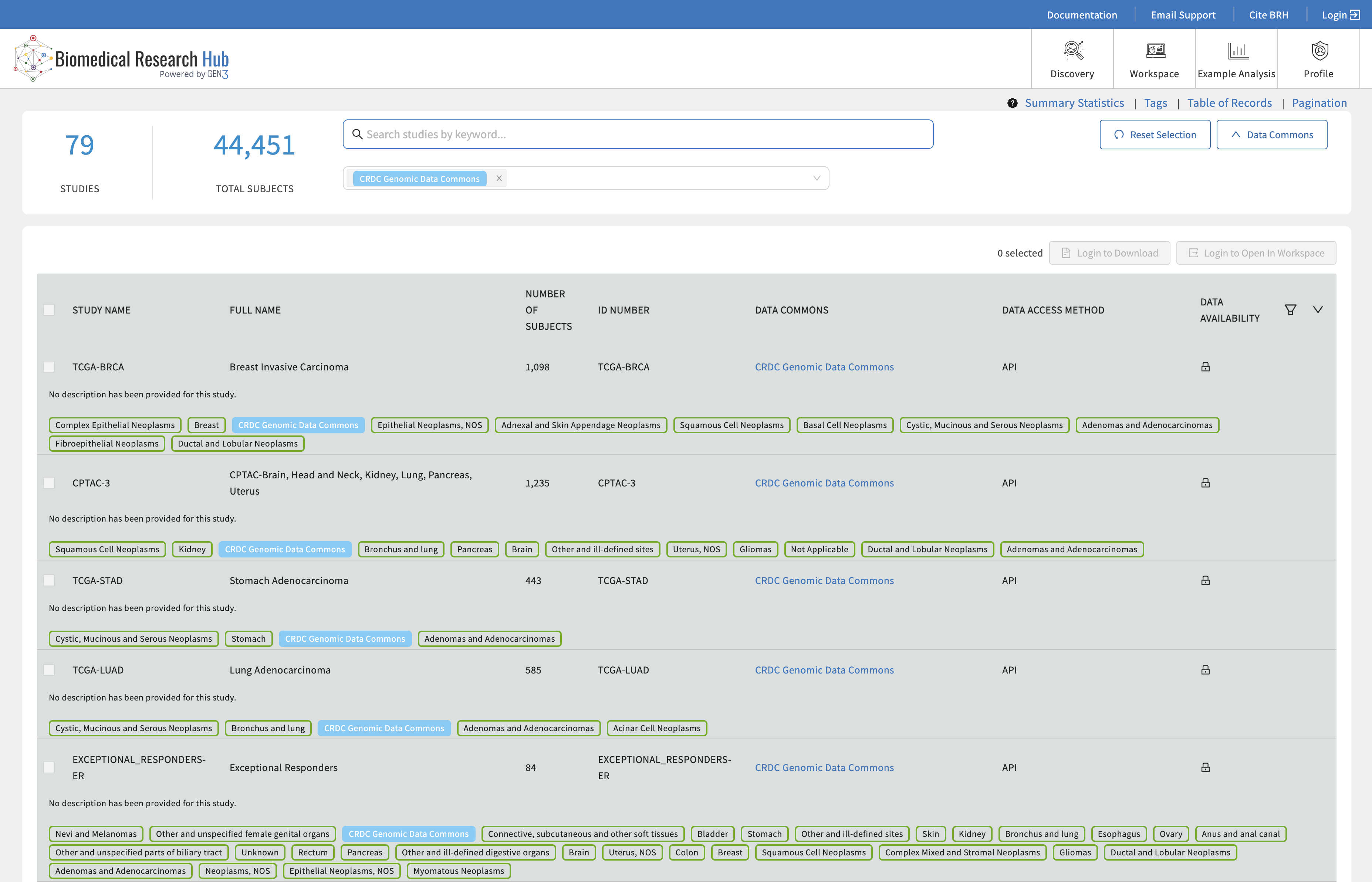}
        \caption{The Discovery page allows users to query the Gen3 metadata service.  For data meshes, like the Biomedical Research Hub shown here, this may be populated with metadata about individual data commons.  It may also be used to contain other semistructured data for an individual data commons such information as about datasets or DOIs.}
    \label{fig:discovery}
\end{figure}

The Exploration page provides the ability to perform faceted searches over structured data managed by a Gen3 data commons.  An alternative is provided by the Discovery page, which provides the ability to search for data objects using the data objects' metadata, which is managed by the MDS.  The data objects can include, for example, datasets, DICOM files, BAM files, or any other data objects managed by a Gen3 data commons.   The discovery page can be configured to pull specifically labeled fields from the MDS and make them searchable.  An example of the Discovery page from the Biomedical Research Hub is shown in Figure \ref{fig:discovery}.

\addcontentsline{toc}{subsection}{Develop Applications using the Gen3 API}
\subsection*{Develop Applications Using the Gen3 API}

Due to the FAIR APIs available for all Gen3 services, it is straightforward to develop applications on top of a Gen3 data commons.  All the necessary documentation for interacting with a particular service can be found in their respective GitHub page within the UC-CDIS organization (https://github.com/uc-cdis).  If an external developer requires changes to a Gen3 service in order to support their use case, there is a process for submitting changes to the open-source Gen3 software (https://docs.gen3.org/gen3-resources/developer-guide/contribute/).

The Pediatric Cancer Data Commons (PCDC) created by Data for the Common Good at the University of Chicago, is a good example of building additional apps and functionality onto the Gen3 software.  The PCDC uses Gen3 for most of its operations, but has built additional functionality in a few distinct areas including more granular authorization, the ability to save filter sets in the data portal, a custom ETL process, and front-end elements including new summary information and a Kaplan-Meier Survival plot.  These improvements are outlined in their recent publication that provides an overview of the PCDC \cite{PCDC}.

Other examples of groups that have created new tools over Gen3 are highlighted on our Gen3 tools page (http://gen3.org/gen3-tools/).  Here you can see a range of tools including those that provide data dictionary visualization, dictionary validation, and ability to work with spreadsheets for data submission.

\addcontentsline{toc}{subsection}{Query a Gen3 Data Commons}
\subsection*{Query a Gen3 Data Commons}

All types of data (structured, semi-structured, and unstructured) are available through the API or the Data Portal.  API queries can be made directly or by using the Gen3 Python SDK tool, which simplifies some of the interactions with the API.  Each type of data has an associated GUID allowing users to have a permanent ID for locating and referencing data.

Structured data follow the data model format created by the commons operators.  Once data are uploaded they are available via graphQL API queries, including within the browser-based GraphiQL query-building interface that is built into the Gen3 data portal. A flattened version of these data is created via ETL to an Elasticsearch index to allow for quicker response times.  The flattened model is queriable via the Exploration page and also via GraphQL API. Structured data are often used to identify which data objects a user may wish to download or explore in a Gen3 Analysis workspace or locally on a user's computer (depending on the size of the file and security restrictions of the commons). Semi-structured data can be queried via API or on a Discovery page.

\addcontentsline{toc}{subsection}{Operating Model}
\subsection*{Operating Model}

To set up, configure and operate a Gen3 data commons requires a range of expertise, including:

\begin{itemize}
    \item {\bf Cloud services.} Someone familiar with the cloud services provided by vendors and systems, such as AWS, GCP, Azure, and OpenStack.
    \item {\bf Software services.} Someone familiar with software and database administration that can set up, customize, and update software applications and manage a database.
    \item {\bf Data services.} Someone familiar with biomedical data who can clean, curate, and harmonize data submitted to the data commons.
\end{itemize}

For smaller commons, this expertise can be provided by a small team of 2-3 people, but a larger-scale data commons typically involves several teams, including the following:

\begin{enumerate}
    \item {\bf Gen3 core developers}.  This includes the Gen3 Core Development Team at the University of Chicago and Gen3 community developers that contribute to the open-source Gen3 core software.
    \item {\bf A cloud services provider (CSP)}.  This may include AWS, GCP, Azure, or a private cloud services provider.
    \item {\bf A commons services provider}. They set up, configure and maintain the Gen3 microservices and other software services required for a Gen3 commons.
    \item {\bf The commons administrator}.  They configure the commons, enhance the user interfaces, set up and update the data model, ingest data, add specialized analysis tools and services, etc.
    \item {\bf Software developers and data scientists}.  They customize the data commons frontend and user interface of the commons, develop specialized analysis tools, integrate the commons with third-party resources, clean and harmonize submitted data,  etc.
    \item {\bf The project sponsor}.  They fund the Gen3 commons and oversee its governance.
\end{enumerate}

\addcontentsline{toc}{subsection}{Gen3 Security}
\subsection*{Gen3 Security}

Gen3 is designed to enable a research group to support secure and compliant data sharing and data analysis, while tailoring the security and compliance policies, procedures, and controls to their specific requirements. Using the terminology of the section above, the different parties' security and compliance responsibilities can be summarized as follows. The Gen3 Core Development team (1) at the University of Chicago, in collaboration with the Gen3 community developers that contribute to the Gen3 core, are responsible for providing stable, secure Gen3 releases and appropriate communications with the open-source community around those releases.
The physical security controls and measures are inherited from the cloud services provider (2).

The Gen3 commons services operator (3) is responsible for understanding all applicable laws and regulations for compliance, Gen3 configurations, continuous integration / continuous deployment (CI/CD) flow, platform operation and monitoring architecture, continuous monitoring (ConMon), policies and procedures related to the specific deployment, and any tools deployed outside of Gen3. The Gen3 common services provider (3) needs a good understanding of security best practices and compliance best practices governing their specific deployment, use case, and data so that they can design a secure and compliant environment with the appropriate configurations. If applicable, the sponsor (6) may define specific use cases or requirements that the commons services operator (3) will configure in Gen3, like an Open ID Connect (OIDC) Identity Provider/System or data use restrictions and policies.

The commons administrator (4) is responsible for managing a particular commons for a specific research community, including setting up and updating the data model, adding community specific analysis tools and services, etc.  The commons administrator (4) must work closely with the commons services provider (3) and commons sponsor (6) to ensure that any changes are compliant with the required security and compliance policies and procedures.

The software developers and data scientists (5) associated with particular commons customize the user interface, develop and integrate specialized analysis tools, clean and harmonize submitted data, etc. From a compliance and security point of view, it is important that any analysis tools that are integrated into the commons or any changes to the user interface maintain all the required security and compliance policies, procedures and controls.

Many of the CTDS-run commons operate at a FedRAMP (Federal Risk and Authorization Management Program) Moderate level.  However, other commons services providers can choose to run a Gen3 system at a higher or lower security standard.

\section*{Data Availability}

This project did not generate any research data, but rather developed open-source software that can be used to manage, analyze and share research data.

\section*{Code Availability}

Gen3 is open source and licensed under Apache License 2.0.  The Gen3 data platform home page \url{https://gen3.org} provides links to the documentation and Github repositories that contain the Gen3 code.  In particular, Gen3 documentation is available from \url{https://docs.gen3.org/} and Gen3 source code is available from \url{https://github.com/uc-cdis}.

\section*{Acknowledgements}

This material is based upon work supported by the National Science Foundation under Grant No. 2346212 and by the  Advanced Research Projects Agency for Health (ARPA-H) under contract 75N92020D00021 / 5N92023F00002.  Any opinions, findings, and conclusions or recommendations expressed in this material are those of the author(s) and do not necessarily reflect the views of the National Science Foundation.  The views and conclusions contained in this document are those of the authors and should not be interpreted as representing the official policies, either expressed or implied, of the U.S. Government.

\section*{Author contributions}
RLG contributed to the conception and high level design of Gen3.
CB, KB, MSF, HPJ, BL, AL, Cl.M, NMS, Ch.M, AP, JQ, RR, PR, LPS, MS, TS, AVT, PV, AV, RLG contributed to the development and operations of Gen3.

\section*{Competing interests}

The authors have no competing interests to declare.
\clearpage
\addcontentsline{toc}{section}{Supplementary Material}
\section*{Supplementary Material}
\clearpage

\begin{figure}
    \includegraphics[width=1.0\textwidth]{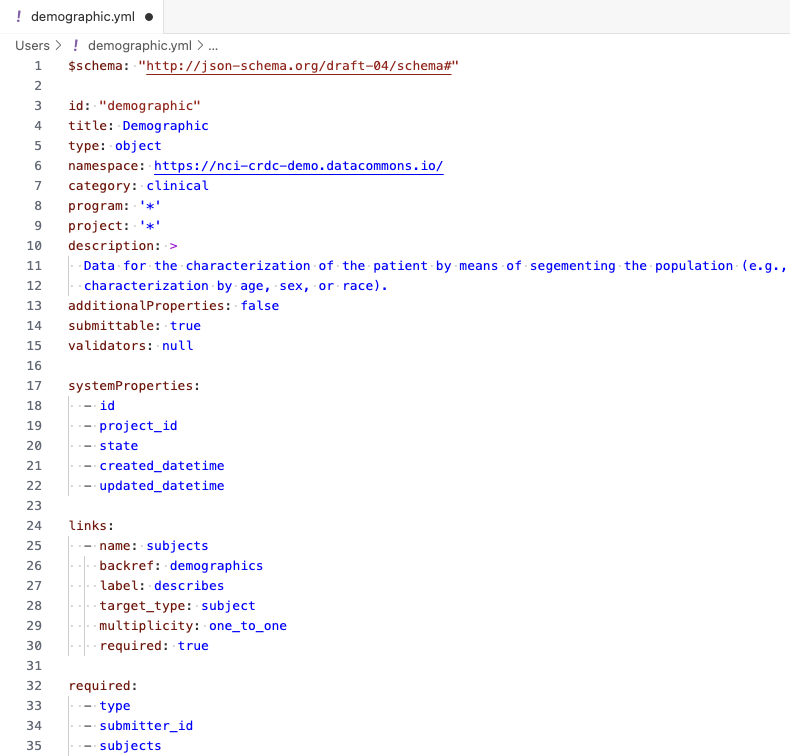}
        \caption{Schema for a node within a Gen3 data model.  This provides a snippet of an example node where properties, permissible values, and links to other nodes are defined. }
    \label{fig:demographic.yaml}
\end{figure}

 \label{table:AWS-services}

\end{document}